**Giant magneto-electric field separation via near-field interference on anapole-like states**

Kseniia Baryshnikova[1,*], Dmitriy Filonov[1,2], Constantin Simovski[1,3], Andrey B. Evlyukhin[1,4], Alexey Kadochkin[1,5], Alaudi Denisultanov[1], Elizaveta Nenasheva[6], Pavel Ginzburg[1,2], and Alexander S. Shalin[1,5,7]

[1] *ITMO University, St. Petersburg 197101, Russia*

[2] *School of Electrical Engineering, Tel Aviv University, Tel Aviv, 69978, Israel*

[3] *Department of Radio Science and Engineering, Aalto University, P.O. Box 13000, FI-00076 Aalto, Finland*

[4] *Laser Zentrum Hannover e.V., Hollerithallee 8, D-30419 Hannover, Germany*

[5] *Ulyanovsk State University, Ulyanovsk, 432017 Russia*

[6] *Giricond Research Institute, Ceramics Co., Ltd., Saint Petersburg 194223, Russia*

[7] *Kotel'nikov Institute of Radio Engineering and Electronics of Russian Academy of Sciences (Ulyanovsk branch), Ulyanovsk, 432071, Russia*

**Abstract**

Quality of spatial separation between electric and magnetic fields in an electromagnetic wave is fundamentally constrained by nonlocal nature of Maxwell's equations. While electric and magnetic energy densities in a wave, propagating in vacuum, are equal at each point in space, carefully designed photonic structures can enable surpassing this limit. Here, a set of high index dielectric tubes was for the first time proposed and theoretically and experimentally demonstrated to deliver a record high spatial separation, overcoming the free space scenario by more than three orders of magnitude with simultaneous enhancement of the magnetic field. Separation effect in the proposed structure is enabled by the near-field interference on anapole-like states, designed by tuning geometrical parameters of coupled dielectric tubes. The void layout of the structure enables the direct observation of the effect with near-field probes and could be further employed for relevant applications. Novel devices, providing tunable high quality separation between electric and magnetic fields, are extremely important for metrology, spectroscopy, spintronics, and opto-electronic applications.

Corresponding author: *k.baryshnikova@optomech.ifmo.ru

A wide range of magnetic phenomena could be treated as relativistic corrections to electrostatic effects[1]. As the result, magnetic contributions are usually orders of magnitude weaker than their electrical counterparts. For example, majority of fundamental light-matter interaction phenomena are driven by electrical dipole transitions, while other multipoles play a minor role[2]. Qualitative analysis shows that the ratio between magnetic and electric dipole transitions in atoms scales with fine-structure constant ($\alpha$), e.g.[3]. Nevertheless, the ability of efficient probing of magnetic lines in fluorescence has a critical role in spectroscopic and metrological applications. Quite a substantial effort was devoted to the investigations of magnetic transitions in atoms and molecules, e.g.[4–6], shining light on their internal structure. One of the major challenges in optical spectroscopic analysis of magnetic contributions is to separate them from other processes that may be orders of magnitude stronger. Enhancement of magnetic and suppression of electric local fields is one of the very promising routes to proceed in order to overcome the above mentioned limitations. This approach, however, demands achieving magneto-electric separations, acceding values of inverse fine-structure constant $\alpha$. Consequently, three-four orders of magnitude difference is a fundamental target to be achieved. A novel auxiliary dielectric structure, proposed here, fits the beforehand mentioned demands.

One of the major challenges in obtaining magneto-optic effects is related to the fact, that natural materials do not exhibit magnetic responses at high frequencies owing to relatively slow spin interactions, defining magnetic susceptibilities. This limitation, however, could be overcome with employing carefully designed nanophotonic structures. For example, strong magnetic responses could be achieved in artificially created media (metamaterials), where specially prepared resonant nanoparticles replicate magnetism in the optical range[7–12]. One of the promising approaches relies on the employment of higher order Mie resonances in high index dielectric particles, e.g. [13–16]. Tuning geometrical parameters of those particles enables achieving many remarkable properties, such as spectral overlapping of multipole resonances[17,18], ultrahigh-quality Fano resonances of scattering[19,20], directive scattering[21-24], or effect of invisibility and cloacking[25,26].

While all the beforehand mentioned approaches enable enhancing local magnetic fields, they usually do not constrained by other conditions, such as suppression of local electric fields. Moreover, inherent coupling between the fields, governed by Maxwell's equations, tends preventing spatial separation between electric and magnetic energies. Efficient magneto-electric separation has numerous applications in metrology and sensing[27,28], fluorescent imaging[29], and many other areas.

The dimensionless magneto-electric separation factor K is defined as:

$$K(\vec{r}) = \eta_0 \frac{|\vec{H}(\vec{r})|}{|\vec{E}(\vec{r})|},\qquad(1)$$

where $\eta_0 = \left|\frac{\vec{E}_0}{\vec{H}_0}\right|$ is the free space wave impedance, while $\vec{H}(\vec{r})$ and $\vec{E}(\vec{r})$ are local magnetic and electric fields correspondingly.

Experimental attempts on magneto-electrical separation have roots back to 1930, where nulling of electrical field were aimed to be achieved at a node of a standing wave, created by a reflection from a mirror[30]. Standing wave interferometric effects were utilized in several applications, such as chirality enhancement[31], Fourier spectrometry[32], and many others. However, practical constraints, such as high precision in mirror positioning, optical alignment, and low tolerance to fabrication inaccuracies, together with fundamentally very steep gradients of the electrical fields at a node, set limitations on this approach. On the other hand, high index dielectric structures open a venue for flexible control over electric and magnetic near-fields and their gradients. However, straightforward approaches on achieving strong magnetic field enhancement should be revised, as the electrical field nulling in majority of the cases is not constrained. In particular, magnetic Mie resonances in high index dielectric particles rely on retardation effects and, as the result, phase accumulation of an incident waves along the particle plays a role. Complete nulling of electrical field at particle's center is usually prevented, if the excitation of higher order resonances (even though they are not under resonant conditions) is not properly addressed. Therefore, careful design of multipole contributions is required with the purpose to achieve destructive interference of electrical local fields. This approach will be undertaken hereafter. Furthermore, in order to meet applications demands, the area with efficient magneto-eclectic separation should be achieved at an externally accessible volume. Therefore, void high index dielectric particles will be consider[13]. In particular, single and coupled dielectric tubes will be investigated analytically, numerically and experimentally. Values of magneto-electric separation factor $K(0) > 5700$, resulting from both local enhancement of the magnetic field and suppression of the electric field will be demonstrated.

The manuscript is organized as follows: a multipole decomposition will be revised first and utilized towards achieving high magneto-electric separation. The next step involves the multipole design in high index dielectric tube-based structures. Since the multipole modes have significant spectral overlap with each other, a minimized, yet sufficient, number of geometrical degrees of freedom is required for achieving high magneto-electric separation. Here, three coupled tubes will be employed and optimized towards this goal. Finally, proof of the concept experiment at GHz spectral range will be demonstrated to support the theoretic predictions.

**Results and Discussion**

*Near-field interference between multipoles*

High index dielectric nanoparticles are known to provide a flexible control over spatial distribution of magnetic fields[11,33,34]. Tubes with coaxial void cores are promising candidates, since the number of geometrical parameters, required for their parameterization, is sufficient for efficient resonant tuning on the one hand, and the void geometry provides the direct access to the local field on the other. Such geometry will be explored hereafter (Fig. 1a). The cartoon shows the schematic nulling of electrical field of the mode, exited by an incident wave, while the magnetic field at the center of the structure is maximal, leading to increased separation coefficient $K(0)$ (or just $K$ hereafter).

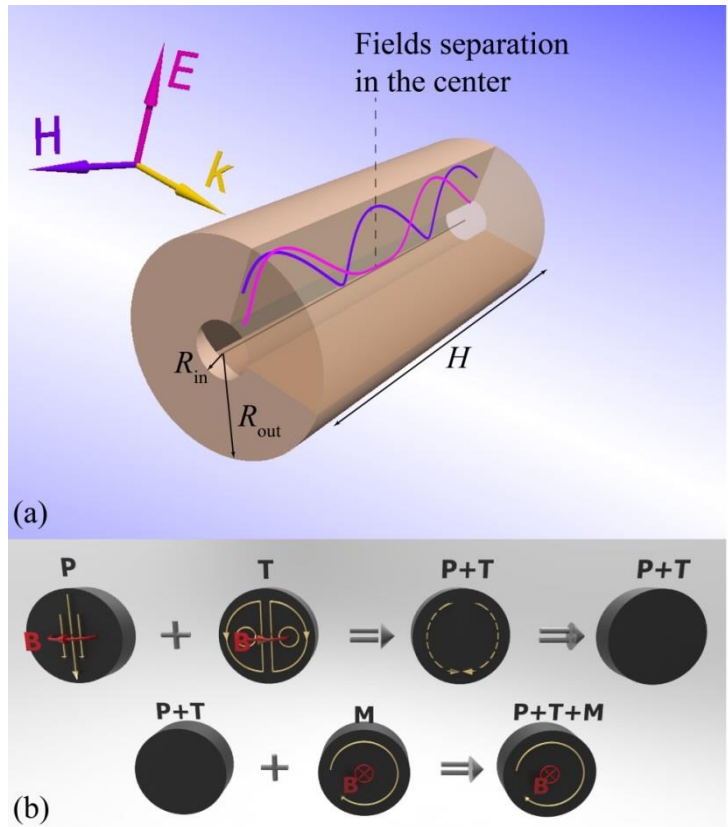

**Figure 1.** Magneto-electric separation concept in all-dielectric structures. (a) Schematics of a dielectric tube, illuminated by a plane wave from a side. Purple and magenta lines represent distributions of magnetic and electric fields along the symmetry axis, respectively. (b) Schematics of multipolar decomposition in a particle – **P**, **T** and **M** stay for electric dipole, electric toroidal dipole and magnetic dipole moments, red lines represent magnetic field **B**, golden lines stay for volumetric displacement currents.

The key effect, responsible for the magneto-electric separation, is the emergence of the anapole state. Since the anapole is not a true eigen mode within the multipole decomposition, but rather the anti-phase combination of dipolar and toroidal moments[25] (Fig. 1b), it will be referred by the name 'state'. The concept of the 'anapole' is further extended here and anapole-like states are defined. This new family of those states demonstrates a further suppression of the dipolar response down to several orders of magnitude owing to additional higher multipoles interference and, as the result, pronounced resonant minimum in the scattering properties can be achieved. Furthermore, those anapole-like states could be designed to have close to zero electrical field at the center of the particle, while magnetic contribution is maximized. The separation effect in the certain points of the system is realized due to the anapole-like state relying on almost complete mutual cancellation of the electric displacement currents, corresponding to the electric dipole and electric toroidal dipole moments (Fig. 1b). If those states are designed to have a proper amplitudes and phases relations electric field in the center will vanish. This is a resonant effect and does not require any conducting materials. At the same time, the magnetic dipole mode could be excited providing magnetic field enhancement in the void (Fig. 1b).

*Proposed all-dielectric structures and numerical results*

Hereafter, the anapole-like states in dielectric tubes will be analyzed numerically and the separation coefficient $K$ at the center of the structure will be maximized. It is worth noting, however, that for the side illumination of the structure (Fig. 1a) the maximal value of $K$ could be obtained at the slightly offset point owing to retardation effects. Nevertheless, geometrical center of the particle will be taken as a reference point for the convenience. Geometrical parameters of the basic building block are - $R_{out}$ and $R_{in}$ are outer and inner radii correspondingly, while $H$ is the height (Fig. 1a). This configuration has a rotational symmetry in respect to the incident field, providing strong toroidal response, as will be shown hereafter.

An isolated dielectric tube was considered first and the magneto-electric separation was maximized by running numerical optimization over all three geometrical parameters. Relatively high separation values of $K$ (as high as 4000) were observed at several geometries, which, however, demand significant aspect ratios $H/2R_{out}$ and $H/2R_{in}$ overcoming values of 10. Being valuable from the theoretical standpoint, those parameters are challenging to achieve with existent fabrication technologies. Remarkably, both optical and microwave realizations possess similar limitations, which, however, originate from different technological aspects (silicon nanoparticles and bulky ceramic devices fabricated in a completely different fashion). On the other hand, high aspect ratios could be obtained with cascading several identical particles on top of each other. Furthermore, in this case, separation distance between adjutant tubes serves as an additional valuable tuning parameter, capable to maximize the magneto-electric separation coefficient and magnetic field inside the void.

The structure, composed of three adjutant tubes (Fig. 2a) is considered next. It is worth noting, that a pair of tubes was tested, but did not show a significant values of separation $K$. The three tubes structure was optimized over all the geometrical parameters, including the separation distance between the tubes $d$. The results for the optimized structure with the following parameters ($R_{out}$ = 198 nm, $H$ = 181 nm, $R_{in}$ = 30 nm; $d$ = 26 nm) appear in Fig. 2. The separation coefficient as high as $K$ = 5730 (and the normalized magnetic field higher than 10) is obtained for the illumination wavelength of 808 nm. This value is within the same order of magnitude to the one reported at[5], where the structure was illuminated with azimuthally polarized beams. Amplitudes of magnetic and electric fields within the three tubes structure appear in Fig. 2b, c. It can be seen, that the region of the maximal magnetic field is at the center of the structure, where electric field is minimized. The field maps have slight asymmetry in respect to the axis of the structure owing to the phase accumulation effect, mentioned before.

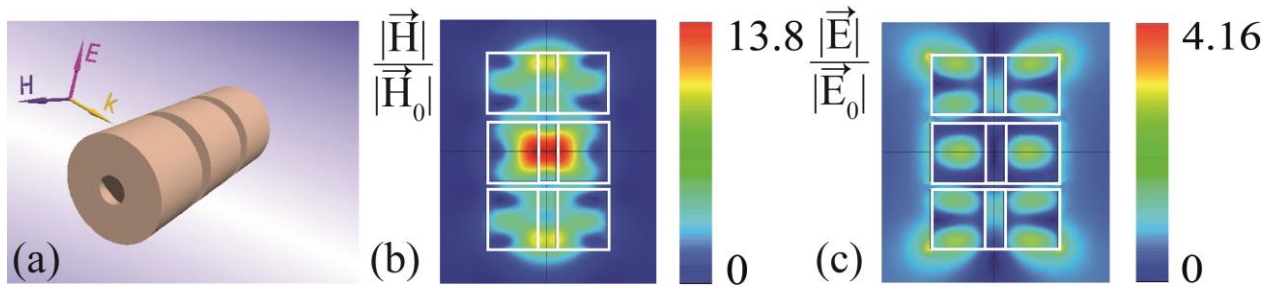

**Figure 2.** Magneto-electric separation in the three tubes system. (a) Schematics of the structure – three closely situated high index dielectric tubes. (b) Magnetic field amplitude enhancement in respect to the free space component of a plane wave. (c) Electric field amplitude enhancement in respect to the free space component of a plane wave. Parameters of the system: $R_{out}$ = 198 nm, $H$ = 181 nm, $R_{in}$ = 30 nm; $d$ = 26 nm, illumination wavelength – 808nm, magneto-electric separation coefficient $K$ = 5730.

In order to verify the anapole-like states contribution to the predicted magneto-electric separation effect, multipolar decomposition of the fields will be performed.

*Multipole decomposition*

Cartesian multipoles and their contributions to the total scattering cross section can be obtained in the irreducible representations[36]. Relevant expressions for calculations of the multipoles and their contributions to the scattering cross section appear at 'Methods'. Figure 3 summarizes the results for the investigated structure. Spectrum of the total electric dipole shows a dip at the same wavelength, where the maximal magneto-electric separation is observed. The total electric dipole response consists of contributions of the electric dipole and electric toroidal dipole moments, having exactly the same scattering pattern in all wave zones[36]. It is worth noting that magnetic dipole resonance and resonant features of others multipoles spectrally shifted from the frequency of maximal separation effect (marked by vertical dashed line in Fig. 3). So one could conclude that anapole-like state is a cause for separation effect that will be proved both numerically and experimentally below.

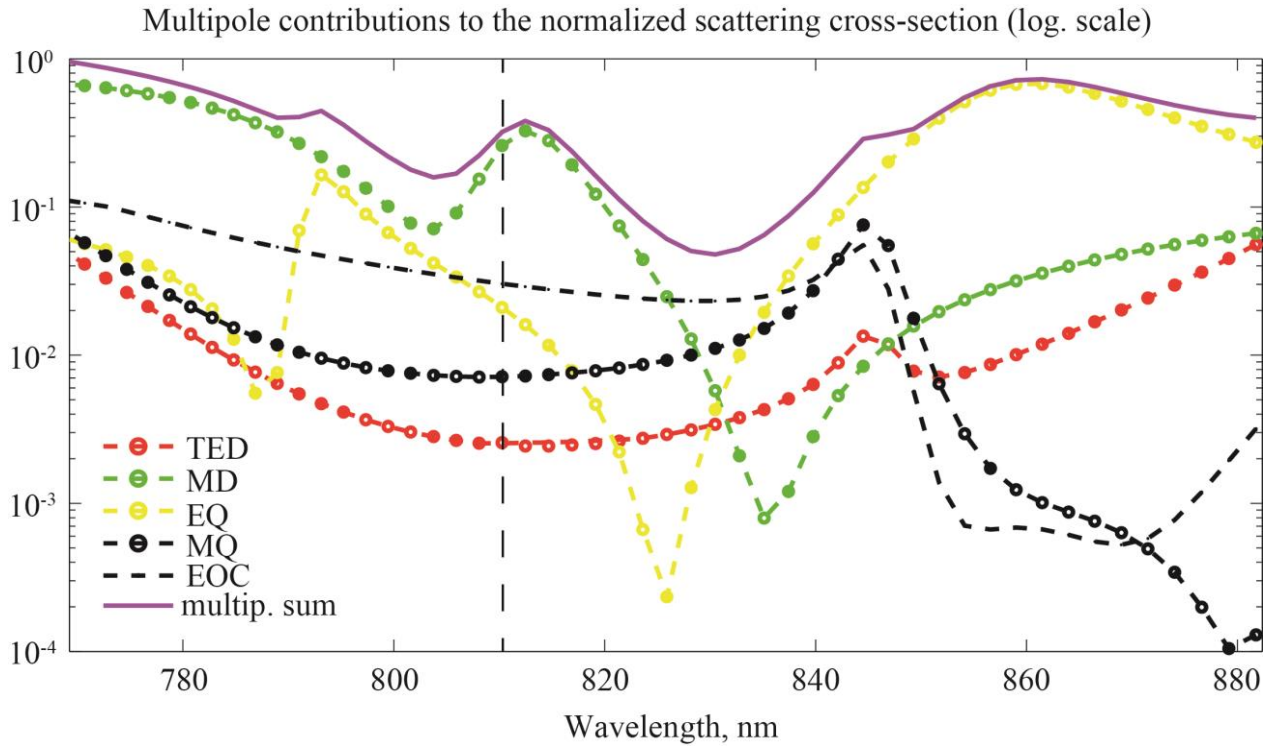

**Figure 3.** Multipole decomposition spectra (logarithmic scale) for the three tubes structure with the parameters $R_{out}$ = 198 nm, $H$ = 181 nm, $R_{in}$ = 30 nm, $d$ = 26 nm, $\lambda$ = 808 nm. Vertical dashed line corresponds to the wavelength of maximal magneto-electric separation with $K$ = 5730. Abbreviations in the legends stay for (TED) – total electric dipole, (MD) – magnetic

dipole, (EQ) – electric quadrupole, (MQ) – magnetic quadrupole, (EOC) - electric octupole, and the purple line stay for the sum of the beforehand mentioned multipoles.

*Experimental results*

Fabrication of nanoscale silicon particles of coaxial shapes could be done with advanced nanolithography techniques. However, experimental realization of a proof-of-concept experiment in the optical domain is very challenging owing to both side excitation and field acquisition schemes that are required. While electric and magnetic near fields, including phases, could be measured with an advanced aparatus[37], direct acquisition of fields inside the structure with near-field probes is an extremely challenging task. On the other hand, mature microwave technology offers numerous tools for advanced analysis of electromagnetic fields. Scalability of Maxwell's equations in respect to the operation frequency enables performing emulation experiments for proof-of-concept demonstrations, e.g.[38] and others. Details of near-field scanning procedure, undertaken at 2 GHz frequency, where the emulation experiment was performed, appear at 'Methods'.

Dielectric tubes were fabricated from $MgO-TiO_2$ ceramics, which has the permittivity of 16, roughly replicating properties of silicon at the visible range. Three identical tubes with, $H = 43.4$ mm, inner radiuses $R_{in} = 8.1$ mm and an outer radiuses $R_{out} = 28.8$ mm were fabricated. The optimal distance between the tubes was found to be 8 mm. Inner and outer radiuses of the tubes are predefined by the fabrication aspects (sizes of the oven templates and polishing machine); hence, the optimization space is reduced. As the result, the maximal magneto-electric separation under those fabrication constrains was found to be 600. Figure 4 summarizes the results of experimental predictions and experimental data. In order to underline the importance of the separation distance, which controls the coupling between the tubes and multipole response of system, two different geometries were compared. The first one (left column of Fig. 4) is the case of touching objects ($d = 0$ mm), while the second one is $d = 8$ mm (optimal design; right column of Fig. 4). It can be seen that the long tube without the gaps (first case) do not show any dip in the total dipole moment spectrum (electric dipole plus electric toroidal dipole), while in the optimized geometry toroidal dipole cancels the dipole mode, giving rise to the anapole state. Numerical and experimental scattering cross section spectra agree well with each other (optical theorem was employed in order to extract the data from the experimental values of the forward scattering) (Fig. 4a, b). Slight differences of experimentally measured and numerically calculated scattering cross-sections are caused by slight variations of samples' geometrical parameters in the range of measurements error. Small physical gap between the touching tubes ($d = 0$ mm) causes additional deviations from the numerical prediction (see Fig. 4a).

Multipole decomposition analysis clearly shows that the strong magnetic dipole resonance predominates other multipoles and dictates the shape of the scattering cross-section in both cases (green circles line in Fig. 4c, d). Increasing the distance between tubes shifts the scattering cross-section peak to lower frequencies. In the case of $d = 8$ mm one can see the strong dip before the resonant frequency. This dip corresponds not only to the minimum of total electric dipole moment (see red circles line in Fig. 4d), but also to the minimum of magnetic dipole moment. Numerical simulations also show that maximal separation coefficient is observed at this frequency. Thus, the separation effect does not solely rely on the magnetic dipole resonance effect, but corresponds to the anapole-like state (the minimum of TED contribution in the scattering), as will be shown hereafter.

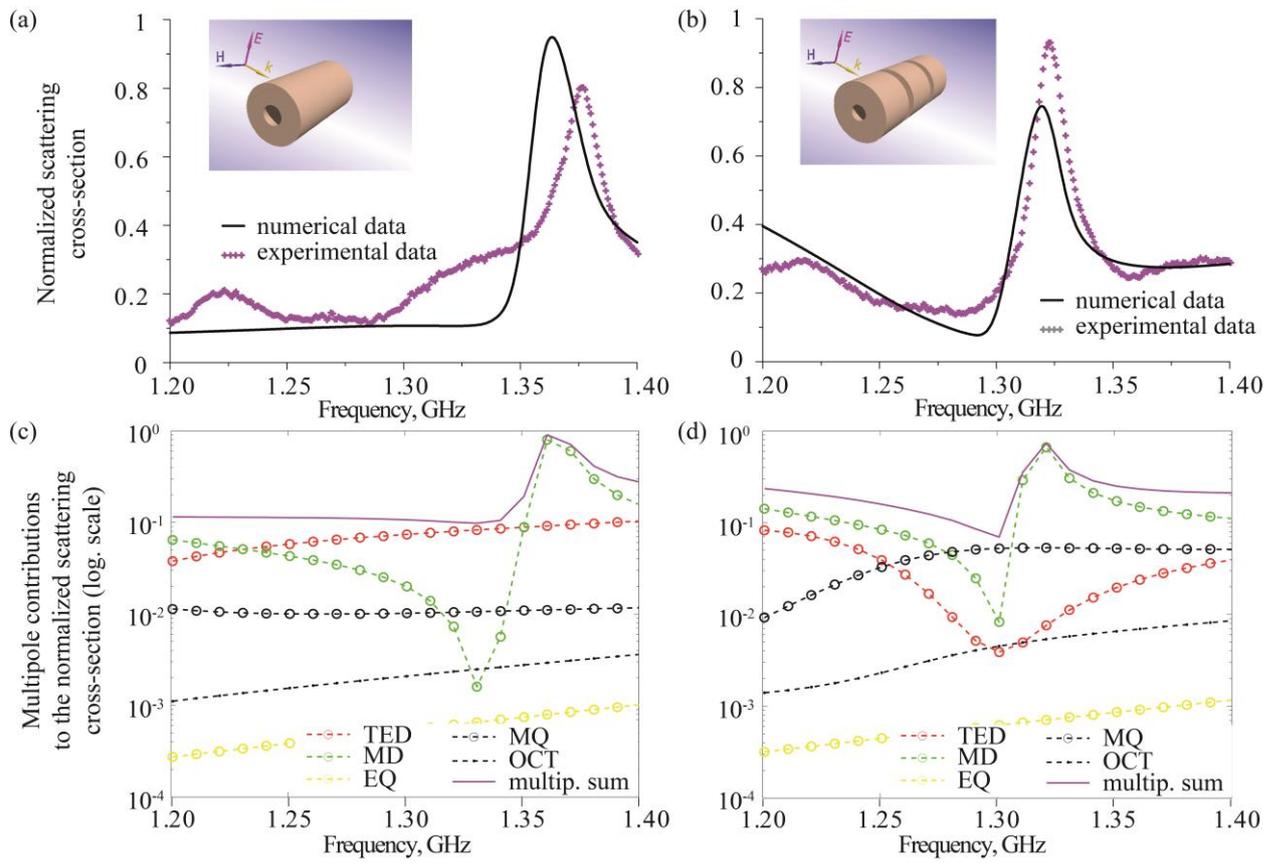

**Figure 4** Magneto-electric separation in ceramic tubes at GHz frequency range. Left column shows results for touching tubes (one solid structure, inset in (a)). Right column shows results for 8 mm distance between the tubes (inset in (b)), for which the maximal magneto-electric separation is achieved. (a, b) Experimental (purple marker line) and numerical (black solid line) normalized scattering cross-section spectra in arbitrary units. (c, d) Multipole decomposition of normalized scattering cross-section spectra, logarithmic scale. Abbreviations are similar to Figure 3.

In order to probe the magneto-separation effect directly, electric and magnetic fields inside the structures were scanned with the near-field probes (see 'Methods' for details). Experimental data shows qualitative agreement with the numerical modeling (Fig. 5), fields in both cases were normalized to the maximum). The results in the touching and optimal separation distance cases are completely different – in the first case the electrical field at the center of the structure does not go down below 30% of the maxima, while in the second case it drops 95% of its initial value, ensuring high magneto-electric separation coefficient (about 600, corresponding to the numerical prediction). Horizontal white lines in Figure 5 highlight the frequency of maximal separation effect. For the case of $d = 0$ mm (Fig 5a, b) numerical simulations show that maximum of magnetic field spectrally shifted far away from the frequency of electric field minimum. Here, the maximal separation correspond to the magnetic dipole resonance frequency and the separation coefficient is modest and below than 100. The case of $d = 8$ mm (Fig. 5c, d) also shows a good qualitative and quantitative agreement between numerical simulations and experimental measurements. In this case maximum of separation correlates with total electric dipole minimum at frequency 1.3 GHz (see white dashed line in Fig. 5c, d), where the separation coefficient achieves values of 600. This result proves the proposed concept of effective magneto-electric separation, relying on anapole-like states.

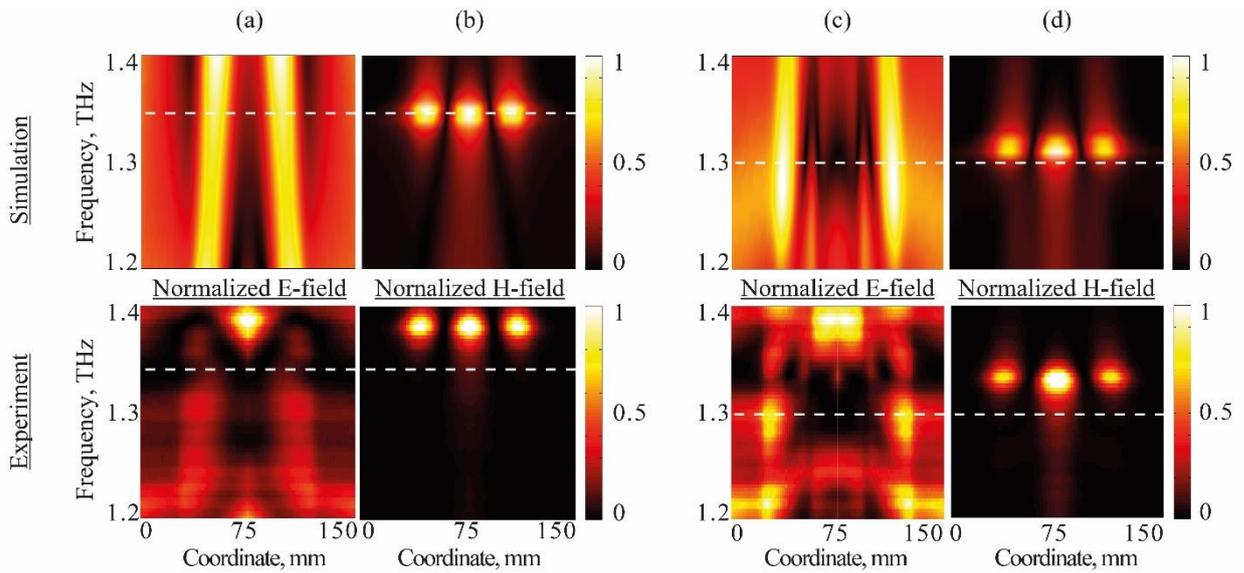

**Figure 5** Normalized electric (a, c) and magnetic (b, d) fields amplitudes on the structure's axis in dependence on coordinate and frequency calculated numerically and measured experimentally. Left column (a, b) shows results for touching tubes. Right column (c, d) shows results for 8 mm distance between the tubes, for which the maximal magneto-electric separation is achieved. Frequency corresponding to the best theoretical separation is marked by the white dashed line.

**Summary and Conclusions**

The concept of magneto-electric field separation in high-index dielectric particles was proposed and analyzed. The core of the effect relies on destructive interference between multipolar contributions. In particular, electric dipole and electric toroidal dipole moments cancel each other, giving rise to anapole family of states, governed by superposition of higher order multipoles. Careful design of electromagnetic multipolar contributions enables achieving superior suppression of electric field in the center of the structure, while the amplitude of the magnetic field is maximized at the same point. A set of three near-field coupled tubes was found to have sufficient number of tuning geometric parameters, and enabled predicting values of magneto-electric separation higher than 5700 with simultaneous more than 10 times enhancement of magnetic field The concept of this optical design was supported with the emulation microwave experiment, where silicon elements were replicated with ceramic objects. Here magneto-electric separation values were predicted to be around 600, which were supported experimentally with a direct near-field scanning.

Electrical field nulling at the center of a void structure could be also utilized for invisibility cover, since an object, placed at the center, does not interact with the incident field[40]. Scattering suppression, being a subject of numerous investigations, e.g.[41,42], could be also achieved with a careful design of multipole interference. Furthermore, efficient magneto-electric separation has numerous applications in metrology, sensing, fluorescent imaging and many other areas.

**Methods**

*Theoretical methods*

Multipole decomposition and scattering cross-section calculations revised in details at[36]. Expressions for leading terms in irreducible Cartesian multipoles representation (up to electric octupole moment) appear below:

Electric dipole moment:

$$\mathbf{p} = \int \mathbf{P}(\mathbf{r}')d\mathbf{r}', \tag{M1}$$

Electric quadrupole moment:

$$\hat{Q} = 3\int \left[ \mathbf{r}'\mathbf{P}(\mathbf{r}') + \mathbf{P}(\mathbf{r}')\mathbf{r}' - \frac{2}{3}(\mathbf{r}' \cdot \mathbf{P}(\mathbf{r}'))\hat{U} \right] d\mathbf{r}', \tag{M2}$$

Magnetic dipole moment:

$$\mathbf{m} = 3\int [\mathbf{r}' \times \mathbf{P}(\mathbf{r}')] d\mathbf{r}', \tag{M3}$$

Magnetic quadrupole moment:

$$\hat{M} = \frac{\omega}{3i} \int \{[\mathbf{r}' \times \mathbf{P}(\mathbf{r}')]\mathbf{r}' + \mathbf{r}'[\mathbf{r}' \times \mathbf{P}(\mathbf{r}')]\} d\mathbf{r}', \tag{M4}$$

Electric octupole moment:

$$O_{\beta\gamma\tau} = (\int \{\mathbf{P}(\mathbf{r}')\mathbf{r}'\mathbf{r}' + \mathbf{r}'\mathbf{P}(\mathbf{r}')\mathbf{r}' + \mathbf{r}'\mathbf{r}'\mathbf{P}(\mathbf{r}')\} d\mathbf{r}')_{\beta\gamma\tau} - (\delta_{\beta\gamma}\Lambda_\tau + \delta_{\beta\tau}\Lambda_\gamma + \delta_{\gamma\tau}\Lambda_\beta) \tag{M5}$$

Electric toroidal dipole moment:

$$\mathbf{T} = \frac{i\omega}{10} \int \{2\mathbf{r}'^2 \mathbf{P}(\mathbf{r}') - (\mathbf{r}' \cdot \mathbf{P}(\mathbf{r}'))\mathbf{r}'\} d\mathbf{r}'. \tag{M6}$$

All integrations are performed over the particle's volume, $\mathbf{r}'$ is the radius-vector, $\mathbf{P}(\mathbf{r}')$ is the polarization vector in the volume, $\delta_{\alpha\beta}$ is the Kronecker delta, and the auxiliary function $\mathbf{\Lambda}$ (used in Eq. M5) is given by:

$$\mathbf{\Lambda} = \frac{1}{5} \int \{2(\mathbf{r}' \cdot \mathbf{P}(\mathbf{r}'))\mathbf{r}' + (\mathbf{r}')^2 \mathbf{P}(\mathbf{r}'))\} d\mathbf{r}'.$$

It is worth noting that this representation implies the explicit separation between electric dipole and electric toroidal dipole moment, that sum up into the total electric dipolar response:

$$\mathbf{D} = \mathbf{p} + \frac{ik_0}{c}\varepsilon_d \mathbf{T}, \tag{M7}$$

where $k_0$ is the wave vector in vacuum, $c$ is the speed of light in vacuum, $\varepsilon_d$ is the dielectric constant of a surrounding environment ($\varepsilon_d = 1$ in the considered case).

The multipole contributions to a scattering cross-section appear below:

Total electric dipole contribution (TED): $$\frac{k_0^4 \sqrt{\varepsilon_d}}{12\pi\varepsilon_0^2 c \mu_0} |\mathbf{D}|^2 , \tag{M8}$$

Electric quadrupole contribution (EQ): $$\frac{k_0^6 \varepsilon_d \sqrt{\varepsilon_d}}{1440\pi\varepsilon_0^2 c \mu_0} \sum |Q_{\alpha\beta}|^2 , \tag{M9}$$

Magnetic dipole contribution (MD): $$\frac{k_0^4 \varepsilon_d \sqrt{\varepsilon_d}}{12\pi\varepsilon_0 c} |\mathbf{m}|^2 , \tag{M10}$$

Magnetic quadrupole contribution (MQ): $$\frac{k_0^6 \varepsilon_d^2 \sqrt{\varepsilon_d}}{160\pi\varepsilon_0 c} \sum |M_{\alpha\beta}|^2 , \tag{M11}$$

Electric octupole contribution (EOC): $$\frac{k_0^8 \varepsilon_d^2 \sqrt{\varepsilon_d}}{3780\pi\varepsilon_0^2 c \mu_0} \sum |O_{\alpha\beta\gamma}|^2 . \tag{M12}$$

Expressions in Eqs. M1-M12 were employed for making all the calculations, reported at the paper.

*Experimental Methods*

Dielectric tubes were fabricated from MgO-TiO$_2$ ceramics, which has the permittivity of 16 (roughly replicating silicon at optical range) and the loss tangent of $(1.12 – 1.17) \times 10^{-4}$ at 2–20 GHz frequency range[43]. Three identical tubes with height $H = 43.4$, inner radiuses $R_{in} = 8.1$ mm and an outer radiuses $R_{out} = 28.8$ mm were fabricated. The optimal distance between the tubes was found to be 8 mm and it was maintained by designing a special holder made of a styrofoam material with dielectric permittivity of 1 at the microwave frequency range. The plane wave excitation was generated by radiating the wave from a rectangular horn antenna connected to the transmitting port of a Vector Network Analyzer (VNA E8362B). The tubes were located at the far-field of the antenna (the distance was 2.5 m), and the second (identical) horn was placed on the optical axis and served as a receiver.

Near-field scanning was performed with an automated step motor device, moving a set of field probes. Small electric dipoles (4 mm length) and shielded loops (2 mm in radius), connected to a rigid coaxial cable, were mounted on the moving holder and fields along the tubes axis were scanned. The cable was connected to the receiving port of the VNA. The probes were oriented in the way they collect the major component of the field, while the connected coaxial cable was kept in the direction perpendicular to the polarization of the incident field (in this case the scattering from the cable is minimal). Direct contact between the probe and the sample was prevented, as was explicitly verified at each scanning point[44].

## References

1. Jackson, J. D. *Classical Electrodynamics*. (John Wiley & Sons; 3rd Edition edition, 1998).

2. Scully, M. O. & Zubairy, M. S. *Quantum Optics*. (Cambridge University Press, 1997).

3. Foot, C. J. *Atomic Physics*. (Oxford University Press, 2005).

4. Deutschbein, O. Experimentelle Untersuchungen über die Vorgänge bei der Licht-emission. *Ann. Phys.* **428,** 183–188 (1939).

5. Kasperczyk, M., Person, S., Ananias, D., Carlos, L. D. & Novotny, L. Excitation of magnetic dipole transitions at optical frequencies. *Phys. Rev. Lett.* **114,** 163903 (2015).

6. Karaveli, S. & Zia, R. Spectral tuning by selective enhancement of electric and magnetic dipole emission. *Phys. Rev. Lett.* **106,** 1–4 (2011).

7. Gomez-Graña, S. *et al.* Hierarchical self-assembly of a bulk metamaterial enables isotropic magnetic permeability at optical frequencies. *Mater. Horiz.* **85,** 3966 (2016).

8. Sheikholeslami, S. N., Alaeian, H., Koh, A. L. & Dionne, J. A. A metafluid exhibiting strong optical magnetism. *Nano Lett.* **13,** 4137–4141 (2013).

9. Qian, Z. *et al.* Raspberry-like metamolecules exhibiting strong magnetic resonances. *ACS Nano* **9,** 1263–1270 (2015).

10. Baryshnikova, K. V, Petrov, M. I., Babicheva, V. E. & Belov, P. A. Plasmonic and silicon spherical nanoparticle anti-reflective coatings. *Sci. Rep.* **6,** 22136 (2016).

11. Baryshnikova, K. V., Novitsky, A., Evlyukhin, A. B. & Shalin, A. S. Magnetic field concentration with coaxial silicon nanocylinders in optical spectral range. *J. Opt. Soc. Am. A* **34,** (2017).

12. Andryieuski, A., Kuznetsova, S. M., & Lavrinenko, A. V. Applicability of point-dipoles approximation to all-dielectric metamaterials. *Physical Review B*, **92**(3), 035114 (2015).

13. Evlyukhin, A. B., Novikov, S. M., Zywietz, U., Eriksen, R. L., Reinhardt, C., Bozhevolnyi, S. I. & Chichkov, B. N. Demonstration of magnetic dipole resonances of dielectric nanospheres in the visible region. *Nano Lett*. **12**, 3749−3755 (2012).

14. Markovich, D. L., Ginzburg, P., Samusev, A. K., Belov, P. A. & Zayats, A. V. Magnetic dipole radiation tailored by substrates: numerical investigation. *Opt. Express* **22,** 10693–10702 (2014).

15. Markovich, D, Baryshnikova, K., Shalin, A., Samusev, A., Krasnok, A., Belov, P., & Ginzburg, P. Enhancement of artificial magnetism via resonant bianisotropy. *Sci. Rep.* **6,** 22546 (2016).

16. Jahani, S., & Jacob, Z. All-dielectric metamaterials. *Nature manotechnology*, **11**(1), 23-36 (2016).

17. Evlyukhin, A. B., Reinhardt, C. & Chichkov, B. N. Multipole light scattering by nonspherical nanoparticles in the discrete dipole approximation. *Phys. Rev. B* **84**, 235429 (2011).

18. Traviss, D. J., Schmidt, M. K., Aizpurua, J., & Muskens, O. L. Antenna resonances in low aspect ratio semiconductor nanowires. *Optics express*, **23**(17), 22771-22787 (2015).

19. Wu, C., *et al*. Spectrally selective chiral silicon metasurfaces based on infrared Fano resonances. *Nature communications*, **5** (2014).

20. Chong, K. E. *et al*. Observation of Fano resonances in all-dielectric nanoparticle oligomers. *Small* **10**(10), 1985-1990 (2014).

21. Liberal, I., Ederra, I., Gonzalo, R. & Ziolkowski, R. W. Superbackscattering from single dielectric particles. *J. Opt.* **17,** 072001 (2015).

22. Terekhov, P. D., Baryshnikova, K. V., Shalin, A. S., Karabchevsky, A. & Evlyukhin, A. B. Resonant forward scattering of light by high-refractive-index dielectric nanoparticles with toroidal dipole contribution. *Opt. Lett.* **42,** 835 (2017).

23. Caldarola, M., *et al*. Non-plasmonic nanoantennas for surface enhanced spectroscopies with ultra-low heat conversion. *Nature communications*, **6** (2015).

24. Krasnok, A. E., Simovski, C. R., Belov, P. A., & Kivshar, Y. S. Superdirective dielectric nanoantennas. *Nanoscale*, **6**(13), 7354-7361 (2014).

25. Miroshnichenko, A. E. *et al*. Nonradiating anapole modes in dielectric nanoparticles. *Nat. Commun.* **6,** 8069 (2015).

26. Mirzaei, A., Miroshnichenko, A. E., Shadrivov, I. V. & Kivshar, Y. S. All-dielectric multilayer cylindrical structures for invisibility cloaking. *Sci. Rep.* **5,** 9574 (2015).

27. Feng, T., Xu, Y., Liang, Z., & Zhang, W. All-dielectric hollow nanodisk for tailoring magnetic dipole emission. *Optics Letters* **41**(21), 5011-5014 (2016).

28. Kasperczyk, M., Person, S., Ananias, D., Carlos, L. D., & Novotny, L. Excitation of magnetic dipole transitions at optical frequencies. *Physical review letters* **114**(16), 163903 (2015).

29. Regmi, R., Berthelot, J., Winkler, P. M., Mivelle, M., Proust, J., Bedu, F., ... & García-Parajó, M. F. All-dielectric silicon nanogap antennas to enhance the fluorescence of single molecules. *Nano Letters* **16**(8), 5143-5151 (2016).

30. Ives, H. E. & Fry, T. C. Standing light waves: repetition of an experiment by Wiener, using a photoelectric probe surface. *J. Opt. Soc. Am.* **23,** 73–83 (1933).

31. Tang, Y. & Cohen, A. E. Enhanced Enantioselectivity in Excitation of Chiral Molecules by Superchiral Light. *Science (80-. ).* **332,** (2011).

32. Le Coarer, E. *et al.* Wavelength-scale stationary-wave integrated Fourier-transform spectrometry. *Nat. Photonics* **1,** 473–478 (2007).

33. Mirzaei, A. & Miroshnichenko, A. E. Electric and magnetic hotspots in dielectric nanowire dimers. *Nanoscale* **7,** 5963–8 (2015).

34. Feng, T., Xu, Y., Liang, Z. & Hang, W. Optically-induced magnetic resonance in all-dielectric hollow nanodisk for tailoring magnetic and electric dipole emission. *Opt. Lett.* **41,** 5011–5014 (2016).

35. Wei, L., Xi, Z., Bhattacharya, N. & Urbach, H. P. Excitation of radiationless anapole mode. *Optica* **1,** 1–4 (2016).

36. Evlyukhin, A. B., Fischer, T., Reinhardt, C. & Chichkov, B. N. Optical theorem and multipole scattering of light by arbitrary shaped nanoparticles. *Phys. Rev. B* **94,** 205434 (2016).

37. Rotenberg, N. & Kuipers, L. Mapping nanoscale light fields. *Nat. Photonics* **8,** 919–926 (2014).

38. Filonov, D., Kramer, Y., Kozlov, V., Malomed, B. a. & Ginzburg, P. Resonant meta-atoms with nonlinearities on demand. *Appl. Phys. Lett.* **109,** (2016).

39. Haar, M. A. Van De, Groep, J. Van De, Brenny, B. J. M. & Polman, A. Controlling magnetic and electric dipole modes in hollow silicon nanocylinders. *Opt. Express* **24,** 191–196 (2016).

40. Alù, A. & Engheta, N. Achieving transparency with plasmonic and metamaterial coatings. *Phys. Rev. E - Stat. Nonlinear, Soft Matter Phys.* **72,** (2005).

41. Filonov, D. S., Shalin, A. S., Iorsh, I., Belov, P. a. & Ginzburg, P. Controlling electromagnetic scattering with wire metamaterial resonators. *J. Opt. Soc. Am. A* **33,** 1910 (2016).

42. Shalin, A. S. *et al.* Scattering suppression from arbitrary objects in spatially dispersive layered metamaterials. *Phys. Rev. B - Condens. Matter Mater. Phys.* **91,** 1–7 (2015).

43. Filonov, D. S. *et al.* Experimental verification of the concept of all-dielectric nanoantennas. *Appl. Phys. Lett.* **100,** 201113 (2012).

44. Filonov, D. S. *et al.* Near-field mapping of Fano resonances in all-dielectric oligomers. *Appl. Phys. Lett.* **104,** 0–4 (2014).